\documentclass[reprint,amsmath,amssymb,aps,floatfix,superscriptaddress,showpacs,prl]{revtex4-1}

\usepackage{graphicx}
\usepackage{dcolumn}
\usepackage{bm}
\usepackage{color}

\begin{document}

\title{Observation of Spectral Structures in the Flux of Cosmic-Ray Protons \\
  from 50 GeV to 60 TeV with the Calorimetric Electron Telescope on International Space Station}

%
\author{O.~Adriani}
\affiliation{Department of Physics, University of Florence, Via Sansone, 1 - 50019, Sesto Fiorentino, Italy}
\affiliation{INFN Sezione di Florence, Via Sansone, 1 - 50019, Sesto Fiorentino, Italy}
\author{Y.~Akaike}
\affiliation{Waseda Research Institute for Science and Engineering, Waseda University, 17 Kikuicho,  Shinjuku, Tokyo 162-0044, Japan}
\affiliation{JEM Utilization Center, Human Spaceflight Technology Directorate, Japan Aerospace Exploration Agency, 2-1-1 Sengen, Tsukuba, Ibaraki 305-8505, Japan}
\author{K.~Asano}
\affiliation{Institute for Cosmic Ray Research, The University of Tokyo, 5-1-5 Kashiwa-no-Ha, Kashiwa, Chiba 277-8582, Japan}
\author{Y.~Asaoka}
\affiliation{Institute for Cosmic Ray Research, The University of Tokyo, 5-1-5 Kashiwa-no-Ha, Kashiwa, Chiba 277-8582, Japan}
\author{E.~Berti} 
\affiliation{Department of Physics, University of Florence, Via Sansone, 1 - 50019, Sesto Fiorentino, Italy}
\affiliation{INFN Sezione di Florence, Via Sansone, 1 - 50019, Sesto Fiorentino, Italy}
\author{G.~Bigongiari}
\affiliation{Department of Physical Sciences, Earth and Environment, University of Siena, via Roma 56, 53100 Siena, Italy}
\affiliation{INFN Sezione di Pisa, Polo Fibonacci, Largo B. Pontecorvo, 3 - 56127 Pisa, Italy}
\author{W.R.~Binns}
\affiliation{Department of Physics and McDonnell Center for the Space Sciences, Washington University, One Brookings Drive, St. Louis, Missouri 63130-4899, USA}
\author{M.~Bongi}
\affiliation{Department of Physics, University of Florence, Via Sansone, 1 - 50019, Sesto Fiorentino, Italy}
\affiliation{INFN Sezione di Florence, Via Sansone, 1 - 50019, Sesto Fiorentino, Italy}
\author{P.~Brogi}
\affiliation{Department of Physical Sciences, Earth and Environment, University of Siena, via Roma 56, 53100 Siena, Italy}
\affiliation{INFN Sezione di Pisa, Polo Fibonacci, Largo B. Pontecorvo, 3 - 56127 Pisa, Italy}
\author{A.~Bruno}
\affiliation{Heliospheric Physics Laboratory, NASA/GSFC, Greenbelt, Maryland 20771, USA}
\author{J.H.~Buckley}
\affiliation{Department of Physics and McDonnell Center for the Space Sciences, Washington University, One Brookings Drive, St. Louis, Missouri 63130-4899, USA}
\author{N.~Cannady}
\affiliation{Center for Space Sciences and Technology, University of Maryland, Baltimore County, 1000 Hilltop Circle, Baltimore, Maryland 21250, USA}
\affiliation{Astroparticle Physics Laboratory, NASA/GSFC, Greenbelt, Maryland 20771, USA}
\affiliation{Center for Research and Exploration in Space Sciences and Technology, NASA/GSFC, Greenbelt, Maryland 20771, USA}
\author{G.~Castellini}
\affiliation{Institute of Applied Physics (IFAC),  National Research Council (CNR), Via Madonna del Piano, 10, 50019, Sesto Fiorentino, Italy}
\author{C.~Checchia}
\affiliation{Department of Physical Sciences, Earth and Environment, University of Siena, via Roma 56, 53100 Siena, Italy}
\affiliation{INFN Sezione di Pisa, Polo Fibonacci, Largo B. Pontecorvo, 3 - 56127 Pisa, Italy}
\author{M.L.~Cherry}
\affiliation{Department of Physics and Astronomy, Louisiana State University, 202 Nicholson Hall, Baton Rouge, Louisiana 70803, USA}
\author{G.~Collazuol}
\affiliation{Department of Physics and Astronomy, University of Padova, Via Marzolo, 8, 35131 Padova, Italy}
\affiliation{INFN Sezione di Padova, Via Marzolo, 8, 35131 Padova, Italy} 
\author{K.~Ebisawa}
\affiliation{Institute of Space and Astronautical Science, Japan Aerospace Exploration Agency, 3-1-1 Yoshinodai, Chuo, Sagamihara, Kanagawa 252-5210, Japan}
\author{A.~W.~Ficklin}
\affiliation{Department of Physics and Astronomy, Louisiana State University, 202 Nicholson Hall, Baton Rouge, Louisiana 70803, USA}
\author{H.~Fuke}
\affiliation{Institute of Space and Astronautical Science, Japan Aerospace Exploration Agency, 3-1-1 Yoshinodai, Chuo, Sagamihara, Kanagawa 252-5210, Japan}
\author{S.~Gonzi}
\affiliation{Department of Physics, University of Florence, Via Sansone, 1 - 50019, Sesto Fiorentino, Italy}
\affiliation{INFN Sezione di Florence, Via Sansone, 1 - 50019, Sesto Fiorentino, Italy}
\author{T.G.~Guzik}
\affiliation{Department of Physics and Astronomy, Louisiana State University, 202 Nicholson Hall, Baton Rouge, Louisiana 70803, USA}
\author{T.~Hams}
\affiliation{Center for Space Sciences and Technology, University of Maryland, Baltimore County, 1000 Hilltop Circle, Baltimore, Maryland 21250, USA}
\author{K.~Hibino}
\affiliation{Kanagawa University, 3-27-1 Rokkakubashi, Kanagawa, Yokohama, Kanagawa 221-8686, Japan}
\author{M.~Ichimura}
\affiliation{Faculty of Science and Technology, Graduate School of Science and Technology, Hirosaki University, 3, Bunkyo, Hirosaki, Aomori 036-8561, Japan}
\author{K.~Ioka}
\affiliation{Yukawa Institute for Theoretical Physics, Kyoto University, Kitashirakawa Oiwake-cho, Sakyo-ku, Kyoto, 606-8502, Japan}
\author{W.~Ishizaki}
\affiliation{Institute for Cosmic Ray Research, The University of Tokyo, 5-1-5 Kashiwa-no-Ha, Kashiwa, Chiba 277-8582, Japan}
\author{M.H.~Israel}
\affiliation{Department of Physics and McDonnell Center for the Space Sciences, Washington University, One Brookings Drive, St. Louis, Missouri 63130-4899, USA}
\author{K.~Kasahara}
\affiliation{Department of Electronic Information Systems, Shibaura Institute of Technology, 307 Fukasaku, Minuma, Saitama 337-8570, Japan}
\author{J.~Kataoka}
\affiliation{School of Advanced Science and	Engineering, Waseda University, 3-4-1 Okubo, Shinjuku, Tokyo 169-8555, Japan}
\author{R.~Kataoka}
\affiliation{National Institute of Polar Research, 10-3, Midori-cho, Tachikawa, Tokyo 190-8518, Japan}
\author{Y.~Katayose}
\affiliation{Faculty of Engineering, Division of Intelligent Systems Engineering, Yokohama National University, 79-5 Tokiwadai, Hodogaya, Yokohama 240-8501, Japan}
\author{C.~Kato}
\affiliation{Faculty of Science, Shinshu University, 3-1-1 Asahi, Matsumoto, Nagano 390-8621, Japan}
\author{N.~Kawanaka}
\affiliation{Yukawa Institute for Theoretical Physics, Kyoto University, Kitashirakawa Oiwake-cho, Sakyo-ku, Kyoto, 606-8502, Japan}
\author{Y.~Kawakubo}
\affiliation{Department of Physics and Astronomy, Louisiana State University, 202 Nicholson Hall, Baton Rouge, Louisiana 70803, USA}
\author{K.~Kobayashi}
\email[]{kenkou@aoni.waseda.jp}
\affiliation{Waseda Research Institute for Science and Engineering, Waseda University, 17 Kikuicho,  Shinjuku, Tokyo 162-0044, Japan}
\affiliation{JEM Utilization Center, Human Spaceflight Technology Directorate, Japan Aerospace Exploration Agency, 2-1-1 Sengen, Tsukuba, Ibaraki 305-8505, Japan}
\author{K.~Kohri} 
\affiliation{Institute of Particle and Nuclear Studies, High Energy Accelerator Research Organization, 1-1 Oho, Tsukuba, Ibaraki, 305-0801, Japan} 
\author{H.S.~Krawczynski}
\affiliation{Department of Physics and McDonnell Center for the Space Sciences, Washington University, One Brookings Drive, St. Louis, Missouri 63130-4899, USA}
\author{J.F.~Krizmanic}
\affiliation{Astroparticle Physics Laboratory, NASA/GSFC, Greenbelt, Maryland 20771, USA}
\author{P.~Maestro}
\affiliation{Department of Physical Sciences, Earth and Environment, University of Siena, via Roma 56, 53100 Siena, Italy}
\affiliation{INFN Sezione di Pisa, Polo Fibonacci, Largo B. Pontecorvo, 3 - 56127 Pisa, Italy}
\author{P.S.~Marrocchesi}
\email[]{marrocchesi@unisi.it}
\affiliation{Department of Physical Sciences, Earth and Environment, University of Siena, via Roma 56, 53100 Siena, Italy}
\affiliation{INFN Sezione di Pisa, Polo Fibonacci, Largo B. Pontecorvo, 3 - 56127 Pisa, Italy}
\author{A.M.~Messineo}
\affiliation{University of Pisa, Polo Fibonacci, Largo B. Pontecorvo, 3 - 56127 Pisa, Italy}
\affiliation{INFN Sezione di Pisa, Polo Fibonacci, Largo B. Pontecorvo, 3 - 56127 Pisa, Italy}
\author{J.W.~Mitchell}
\affiliation{Astroparticle Physics Laboratory, NASA/GSFC, Greenbelt, Maryland 20771, USA}
\author{S.~Miyake}
\affiliation{Department of Electrical and Electronic Systems Engineering, National Institute of Technology (KOSEN), Ibaraki College, 866 Nakane, Hitachinaka, Ibaraki 312-8508, Japan}
\author{A.A.~Moiseev}
\affiliation{Department of Astronomy, University of Maryland, College Park, Maryland 20742, USA}
\affiliation{Astroparticle Physics Laboratory, NASA/GSFC, Greenbelt, Maryland 20771, USA}
\affiliation{Center for Research and Exploration in Space Sciences and Technology, NASA/GSFC, Greenbelt, Maryland 20771, USA}
\author{M.~Mori}
\affiliation{Department of Physical Sciences, College of Science and Engineering, Ritsumeikan University, Shiga 525-8577, Japan}
\author{N.~Mori}
\affiliation{INFN Sezione di Florence, Via Sansone, 1 - 50019, Sesto Fiorentino, Italy}
\author{H.M.~Motz}
\affiliation{Faculty of Science and Engineering, Global Center for Science and Engineering, Waseda University, 3-4-1 Okubo, Shinjuku, Tokyo 169-8555, Japan}
\author{K.~Munakata}
\affiliation{Faculty of Science, Shinshu University, 3-1-1 Asahi, Matsumoto, Nagano 390-8621, Japan}
\author{S.~Nakahira}
\affiliation{Institute of Space and Astronautical Science, Japan Aerospace Exploration Agency, 3-1-1 Yoshinodai, Chuo, Sagamihara, Kanagawa 252-5210, Japan}
\author{J.~Nishimura}
\affiliation{Institute of Space and Astronautical Science, Japan Aerospace Exploration Agency, 3-1-1 Yoshinodai, Chuo, Sagamihara, Kanagawa 252-5210, Japan}
\author{G.A.~de~Nolfo}
\affiliation{Heliospheric Physics Laboratory, NASA/GSFC, Greenbelt, Maryland 20771, USA}
\author{S.~Okuno}
\affiliation{Kanagawa University, 3-27-1 Rokkakubashi, Kanagawa, Yokohama, Kanagawa 221-8686, Japan}
\author{J.F.~Ormes}
\affiliation{Department of Physics and Astronomy, University of Denver, Physics Building, Room 211, 2112 East Wesley Avenue, Denver, Colorado 80208-6900, USA}
\author{S.~Ozawa}
\affiliation{Quantum ICT Advanced Development Center, National Institute of Information and Communications Technology, 4-2-1 Nukui-Kitamachi, Koganei, Tokyo 184-8795, Japan}
\author{L.~Pacini}
\affiliation{Department of Physics, University of Florence, Via Sansone, 1 - 50019, Sesto Fiorentino, Italy}
\affiliation{Institute of Applied Physics (IFAC),  National Research Council (CNR), Via Madonna del Piano, 10, 50019, Sesto Fiorentino, Italy}
\affiliation{INFN Sezione di Florence, Via Sansone, 1 - 50019, Sesto Fiorentino, Italy}
\author{P.~Papini}
\affiliation{INFN Sezione di Florence, Via Sansone, 1 - 50019, Sesto Fiorentino, Italy}
\author{B.F.~Rauch}
\affiliation{Department of Physics and McDonnell Center for the Space Sciences, Washington University, One Brookings Drive, St. Louis, Missouri 63130-4899, USA}
\author{S.B.~Ricciarini}
\affiliation{Institute of Applied Physics (IFAC),  National Research Council (CNR), Via Madonna del Piano, 10, 50019, Sesto Fiorentino, Italy}
\affiliation{INFN Sezione di Florence, Via Sansone, 1 - 50019, Sesto Fiorentino, Italy}
\author{K.~Sakai}
\affiliation{Center for Space Sciences and Technology, University of Maryland, Baltimore County, 1000 Hilltop Circle, Baltimore, Maryland 21250, USA}
\affiliation{Astroparticle Physics Laboratory, NASA/GSFC, Greenbelt, Maryland 20771, USA}
\affiliation{Center for Research and Exploration in Space Sciences and Technology, NASA/GSFC, Greenbelt, Maryland 20771, USA}
\author{T.~Sakamoto}
\affiliation{College of Science and Engineering, Department of Physics and Mathematics, Aoyama Gakuin University,  5-10-1 Fuchinobe, Chuo, Sagamihara, Kanagawa 252-5258, Japan}
\author{M.~Sasaki}
\affiliation{Department of Astronomy, University of Maryland, College Park, Maryland 20742, USA}
\affiliation{Astroparticle Physics Laboratory, NASA/GSFC, Greenbelt, Maryland 20771, USA}
\affiliation{Center for Research and Exploration in Space Sciences and Technology, NASA/GSFC, Greenbelt, Maryland 20771, USA}
\author{Y.~Shimizu}
\affiliation{Kanagawa University, 3-27-1 Rokkakubashi, Kanagawa, Yokohama, Kanagawa 221-8686, Japan}
\author{A.~Shiomi}
\affiliation{College of Industrial Technology, Nihon University, 1-2-1 Izumi, Narashino, Chiba 275-8575, Japan}
\author{P.~Spillantini}
\affiliation{Department of Physics, University of Florence, Via Sansone, 1 - 50019, Sesto Fiorentino, Italy}
\author{F.~Stolzi}
\affiliation{Department of Physical Sciences, Earth and Environment, University of Siena, via Roma 56, 53100 Siena, Italy}
\affiliation{INFN Sezione di Pisa, Polo Fibonacci, Largo B. Pontecorvo, 3 - 56127 Pisa, Italy}
\author{S.~Sugita}
\affiliation{College of Science and Engineering, Department of Physics and Mathematics, Aoyama Gakuin University,  5-10-1 Fuchinobe, Chuo, Sagamihara, Kanagawa 252-5258, Japan}
\author{A.~Sulaj} 
\affiliation{Department of Physical Sciences, Earth and Environment, University of Siena, via Roma 56, 53100 Siena, Italy}
\affiliation{INFN Sezione di Pisa, Polo Fibonacci, Largo B. Pontecorvo, 3 - 56127 Pisa, Italy}
\author{M.~Takita}
\affiliation{Institute for Cosmic Ray Research, The University of Tokyo, 5-1-5 Kashiwa-no-Ha, Kashiwa, Chiba 277-8582, Japan}
\author{T.~Tamura}
\affiliation{Kanagawa University, 3-27-1 Rokkakubashi, Kanagawa, Yokohama, Kanagawa 221-8686, Japan}
\author{T.~Terasawa}
\affiliation{Institute for Cosmic Ray Research, The University of Tokyo, 5-1-5 Kashiwa-no-Ha, Kashiwa, Chiba 277-8582, Japan}
\author{S.~Torii}
\email[]{torii.shoji@waseda.jp}
\affiliation{Waseda Research Institute for Science and Engineering, Waseda University, 17 Kikuicho,  Shinjuku, Tokyo 162-0044, Japan}
\author{Y.~Tsunesada}
\affiliation{Graduate School of Science, Osaka Metropolitan University, Sugimoto, Sumiyoshi, Osaka 558-8585, Japan }
\affiliation{ Nambu Yoichiro Institute for Theoretical and Experimental Physics, Osaka Metropolitan University,  Sugimoto, Sumiyoshi, Osaka  558-8585, Japan}
\author{Y.~Uchihori}
\affiliation{National Institutes for Quantum and Radiation Science and Technology, 4-9-1 Anagawa, Inage, Chiba 263-8555, Japan}
\author{E.~Vannuccini}
\affiliation{INFN Sezione di Florence, Via Sansone, 1 - 50019, Sesto Fiorentino, Italy}
\author{J.P.~Wefel}
\affiliation{Department of Physics and Astronomy, Louisiana State University, 202 Nicholson Hall, Baton Rouge, Louisiana 70803, USA}
\author{K.~Yamaoka}
\affiliation{Nagoya University, Furo, Chikusa, Nagoya 464-8601, Japan}
\author{S.~Yanagita}
\affiliation{College of Science, Ibaraki University, 2-1-1 Bunkyo, Mito, Ibaraki 310-8512, Japan}
\author{A.~Yoshida}
\affiliation{College of Science and Engineering, Department of Physics and Mathematics, Aoyama Gakuin University,  5-10-1 Fuchinobe, Chuo, Sagamihara, Kanagawa 252-5258, Japan}
\author{K.~Yoshida}
\affiliation{Department of Electronic Information Systems, Shibaura Institute of Technology, 307 Fukasaku, Minuma, Saitama 337-8570, Japan}
\author{W.~V.~Zober}
\affiliation{Department of Physics and McDonnell Center for the Space Sciences, Washington University, One Brookings Drive, St. Louis, Missouri 63130-4899, USA}

\collaboration{CALET Collaboration}

\date{\today}

\begin{abstract}
  A precise measurement of the cosmic-ray proton spectrum with the Calorimetric Electron Telescope (CALET) is presented in the energy interval from 50~GeV to 60~TeV and the observation of a softening of the spectrum above 10 TeV is reported. The analysis is based on the data collected during $\sim$6.2 years of smooth operations aboard the International Space Station (ISS) 
  and covers a broader energy range with respect to the previous proton flux measurement by CALET, 
  with an increase of the available statistics by a factor of $\sim$2.2.
 Above a few hundred GeV we confirm our previous observation of a progressive spectral hardening with a higher significance (more than 20 sigma). 
 In the multi-TeV region 
 we observe a second spectral feature with a softening around 10 TeV
 and a spectral index change from $-2.6$ to $-2.9$ consistently, within the errors, with the shape of the spectrum reported by DAMPE.
  We apply a simultaneous fit of the proton differential spectrum which well reproduces the gradual change of the spectral index encompassing the lower energy power-law regime and the two spectral features observed at higher energies. 
\end{abstract}


\maketitle

$\bf{\it Introduction.-}$Recent direct measurements of cosmic rays have shown the presence of unexpected spectral    structures significantly departing from a simple-power-law dependence.
The presence of a spectral hardening has been established for several nuclear species~\cite{ATIC2-p,CREAM-nuclei,CREAM-hardening,CREAM-I,PAMELA,AMS-02-proton,AMS-02-helium,CREAM-III-pHe,AMS-02-carbon,AMS-02-boron,AMS-02-nitrogen, 
CALET-p, CALET-CO} around a few hundred GeV/${n}$  
and high statistics measurements have shown that the rigidity dependence of primary and secondary cosmic nuclei is different \cite{AMS02_NeMgSi}.\\
\indent
This rich phenomenology has been addressed by several theoretical models
in the quest for a consistent picture of cosmic-ray acceleration (eventually including new sources)~\cite{IP1-Ellison-1997, IP2-Malkov-2012, LS1-Erlykin-2012, SB1-Drury-2011, SB3-Ohira-2016, SS1-Biermann-2010, SS2-Ptuskin-2013, SS3-Zatsepin-2006,  Vladmirov-2012, kawanaka-2018}, propagation (or reacceleration) in the Galaxy ~\cite{PR1-Blasi-2012, PR2-Aloisio-2013, RA2-Thoudam, Tomassetti-2012, Tomassetti2015, Semikoz2018, Evoli2018} and the possible presence of one or more local sources~\cite{LS2-Thoudam-2012, LS3-Bernard-2013}. 
More recent theoretical contributions were presented at the International Cosmic Ray 2021 Conference~\cite{Lipari-ICRC21, Recchia-ICRC21, Caprioli-ICRC21, Cristofari-ICRC21, Morlino-ICRC21}. %
The hypothesis of a possible charge-dependent cutoff in the nuclei spectra can be directly tested with long duration measurements in space, provided they achieve a sufficient exposure, adequate energy resolution, and the capability to identify individual elements.\\
\indent
New data from space-borne calorimetric instruments have recently become available, expanding the energy frontier of proton measurements by more than 1 order of magnitude.  Following our previous observation up to 10 TeV of a spectral hardening of the proton spectrum
around a few hundred GeV,
a new feature emerged above 10 TeV whereby the spectral index was found to gradually change and a softening of the spectrum was clearly observed, as also reported by DAMPE~\cite{DAMPE_proton} and CALET~\cite{ICRC2021_KK&PSM} and previously by NUCLEON~\cite{NUCLEON-JTEP}
 and CREAM-III~\cite{CREAM3_proton}.\\
\indent
For proton and helium, it is important to determine the detailed rigidity dependence of the spectral index through the whole spectrum, studying the onset of the spectral hardening and of the softening regime at higher energy, respectively.
In order to achieve a consistent picture, systematic errors should be kept under control and a critical comparison of the observations from different experiments should be fostered.\\
\indent
The Calorimetric Electron Telescope (CALET)
~\cite{Pier-ICRC21, asaoka2018}, in operation on the International Space Station since 2015, is
a calorimetric instrument optimized for the measurement of the all-electron spectrum~\cite{CALET2017,CALET2018}. 
It has enough depth, dynamic range, and energy resolution to
measure protons, helium \cite{ICRC2021_helium}, and heavier cosmic-ray nuclei
(up to iron and above)~\cite{CALET-CO, ICRC2021_carbon,  ICRC2021_boron, CALET-IRON2021, ICRC2021_iron, CALET-Nickel22, ICRC2021_Zober} at energies reaching the PeV scale. \\
\indent
In this Letter, we present a direct measurement of the cosmic-ray proton differential spectrum in kinetic energy from 50~GeV to 60~TeV with CALET.\\
\newline
\indent
$\bf{\it The~CALET~instrument.-}$Designed to achieve a full containment of TeV electromagnetic showers and a large
 electron-proton discrimination capability ($>$10$^{5}$), it is longitudinally segmented into a fine grained imaging calorimeter (IMC) followed by a total absorption calorimeter (TASC). The TASC is a 27 $X_0$ (radiation length) thick homogeneous calorimeter with 12 alternate orthogonal layers of lead-tungstate logs. 
The IMC is a sampling calorimeter segmented into 16 layers of individually read-out scintillating fibers (with 1 mm$^2$ square cross section)  and interspaced with thin tungsten absorbers. Alternate planes of fibers are arranged along orthogonal directions.
It can image the early shower profile in the first 3 $X_0$ and provide tracking information by reconstructing the incident direction of cosmic rays with good angular resolution (0.1$^\circ$ for electrons and better than 0.5$^\circ$ for hadrons) \cite{Torii-ICRC19}.
The overall thickness of CALET at normal incidence is 30 $X_0$ and $\sim$1.3 $\lambda_{I}$ (proton interaction lengths).
The charge identification of individual nuclear species is performed by a two-layered hodoscope of plastic scintillators (CHD), positioned at the top of the apparatus, providing a measurement of the charge $Z$ of the incident particle over a wide dynamic range ($Z=1$ to $\sim40$) with sufficient charge resolution to resolve individual elements \cite{Mar2011} and complemented by a redundant charge determination via multiple $dE/dx$ measurements in the IMC. 
The overall CHD charge resolution (in $Z$ units)
increases linearly, as a function of the atomic number, from $\sim$0.1 for protons to $\sim$0.3 for iron.  For the IMC, 
multiple sampling in the IMC achieves an excellent performance as shown in
 Ref.~\cite{Pier-ICRC19} where the charge resolution is plotted as a function of the atomic number $Z$.
The interaction point (IP) is first reconstructed \cite{Brogi-ICRC15}, and only the $dE/dx$ ionization clusters from the layers upstream of the IP are used to infer a charge value from the truncated mean of the valid samples.
The geometrical factor of CALET is $\sim$0.1 m$^2$sr, and the total weight is 613 kg.  The instrument is described in more detail elsewhere \cite{asaoka2017}.\\
\newline
\indent
$\bf{\it Data~Analysis.-}$Flight data collected for 2272 days from October 13, 2015, to December 31, 2021, were analyzed.
The total observation live time with the high-energy (HE) shower trigger~\cite{asaoka2018}
 is 1925 days. A low-energy (LE) shower trigger,
 operated at high geomagnetic latitudes~\cite{asaoka2018}, was also used for the analysis of the
 low-energy region. As we have sufficient statistics for protons below 100 GeV,
 we used the data presented in Ref.~\cite{CALET-p}.

A Monte Carlo (MC) simulation, based on the EPICS simulation package \cite{EPICS,EPICSver},
was developed to reproduce the detailed detector configuration and physics processes, as well as detector signals.
In order to assess the uncertainties due to the modeling of hadronic interactions,
a series of beam tests were carried out at the CERN-SPS with proton beams of 30, 100, and 400 GeV. 
However, no beam test calibrations are possible beyond this limit with the available accelerated beams. Therefore simulations with FLUKA~\cite{FLUKA-1,FLUKA-2,FLUKAver} and GEANT4~\cite{Geant4,Geant4ver} were compared with 
 EPICS, and the differences were properly accounted for in the systematic uncertainties. Trigger efficiency and energy response derived from MC simulations were extensively studied
 ~\cite{CALET-p}.

As described in our previous publication~\cite{CALET-p}, the track of the primary cosmic-ray particle was reconstructed from the hit pattern of the IMC fibers 
by means of a Kalman filter tracking package~\cite{paolo2017} developed for CALET. The shower energy is calculated as the sum of the TASC energy deposits. The total observed energy ($E_\text{TASC}$)
is calibrated using penetrating particles, and
 a seamless stitching of adjacent gain ranges is performed on orbit. This procedure was complemented by the confirmation of the linearity
of the system over the whole range by means of ground
measurements using a UV pulsed laser, as described in Ref.~\cite{asaoka2017}. Temporal variations during the long-term observation period
were also corrected for, using penetrating
particles to monitor the gain of each sensor~\cite{CALET2017}.

In order to minimize the background contamination, the following 
criteria were applied to well-reconstructed and well-contained
proton-events:
(1) off-line trigger confirmation,
(2) geometrical acceptance condition (requires acceptance type $A$ as defined in Ref.~\cite{CALET2018}),
(3) reliability of the reconstructed track while retaining a high efficiency,
(4) electron rejection,
(5) rejection of off-acceptance events,
(6) consistency of the track impact point in the TASC with the calorimetric energy deposits,
(7) requirements on the shower development in the IMC, and
(8) identification of the particle as a proton by using both CHD and IMC charge measurements.

Criterion (1) applies more stringent conditions with respect to the onboard trigger 
removing effects caused by positional and temporal variations of the detector gain.
In the energy range $E>300$ GeV, the HE trigger should be asserted and the energy deposit sum
of the IMC 7th and 8th layers is required to exceed 50 minimum ionizing particles (MIPs) in either the
$X$ or $Y$ view. Furthermore, the energy deposit of the first TASC layer (TASC-X1) should be larger than 100 MIPs.
In the energy range $E<300$ GeV, the LE trigger should be asserted, the energy deposit sum
of the IMC layers 7 and 8 should be greater than 5 MIPs in either the $X$ or $Y$ view and the energy
deposit of TASC-X1 should be larger than 10 MIPs.
Criterion (3) requires the reliability of track fitting (details on track quality cuts can be found in the Supplemental Material of Ref.~\cite{CALET-p}). 

In order to reject electron events, a ``Moli\`{e}re concentration'' along the track
 is calculated by summing up all energy deposits observed inside one Moli\`{e}re
radius for tungsten ($\pm$9 fibers, i.e., 9 mm) around the IMC fiber 
best matched with the track.  By requiring the ratio of the energy deposit within 9 mm to the total energy
deposit sum in the IMC to be
less than 0.7 [criterion (4)], most of the electrons are rejected while retaining
an efficiency above 92\% for protons.

In order to minimize the fraction of misidentified events, two topological
cuts are applied using the TASC energy-deposit information only and
irrespective of IMC tracking [criterion (5)].
These cuts remove poorly reconstructed events where one of the
secondary tracks is identified as the primary track (refer to the Supplemental Material of Ref.~\cite{CALET-p}).

Criterion (6) removes additional misreconstructed events by applying a consistency cut
between the track impact point and the center of gravity of
the energy deposits in the first and second (TASC-Y1)
layers of the TASC.

In order to select well-contained events, energy dependent thresholds are
set to achieve a 95\% constant efficiency for events
that interacted in the IMC below the fourth layer [criterion (7)].
After applying criteria (1) -- (7), 
charge, energy, and trigger efficiency are determined for the selected sample
(hereafter denoted as ``target'' events).

Backscattered particles from the calorimeter can affect both
the trigger and the charge determination. In fact, 
primary particles below the trigger thresholds might be 
triggered anyway because of backscattered particles hitting 
the TASC-X1 and IMC bottom layers. 
Moreover a significant amount of  
backscatter may potentially induce a fake charge identification
by increasing the number of hits with a significant energy deposit in IMC and CHD [criterion (8)].

The charge $Z$ is calculated as
 $Z = a(E) N_\text{mip}^{b(E)/2}$,
where $N_\text{mip}$ is the CHD or IMC response (in MIP units) and $a(E)$ and $b(E)$ are energy dependent charge correction coefficients (mainly accounting for backscattering effects increasing with energy) applied separately to flight data, EPICS, FLUKA, and GEANT4 to optimize the determination of the charge peaks of proton and helium at $Z=1$ and 2, respectively \cite{CALET-p}.
 
\begin{figure}[bt!] 
  \begin{center}  
    \includegraphics[width=7cm]{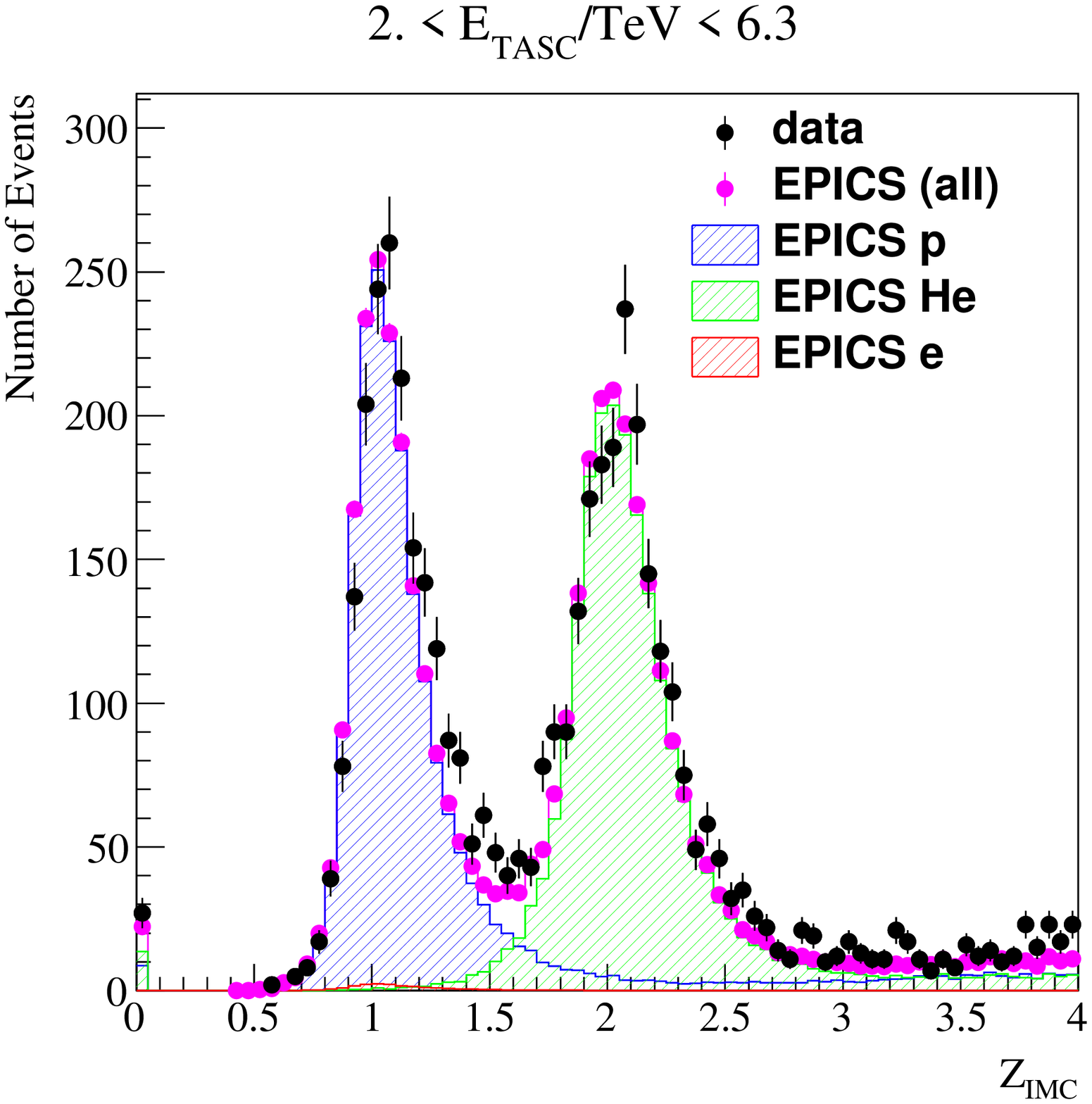}
    \includegraphics[width=7cm]{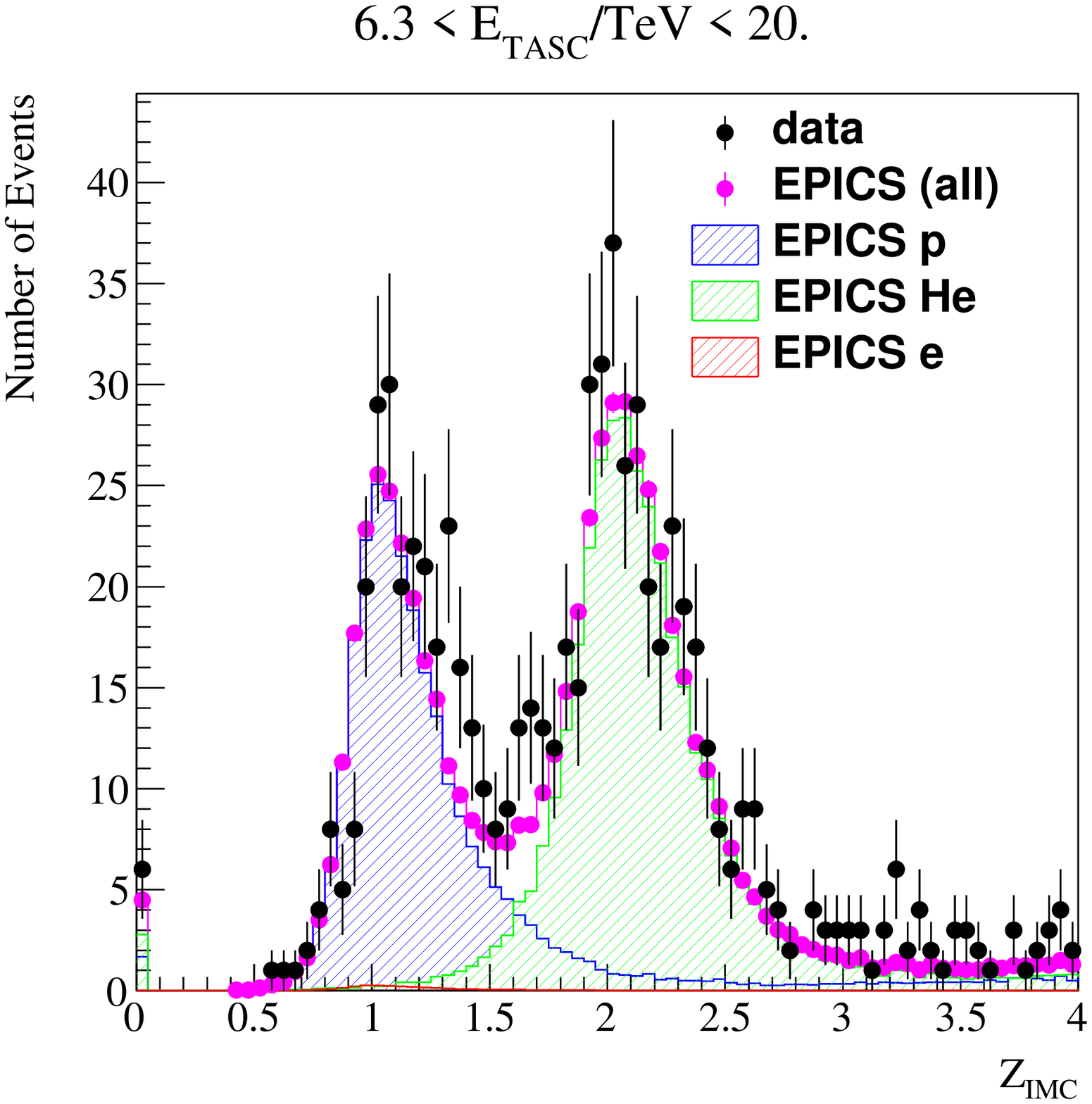}
    \caption{Examples of charge distributions with the IMC compared with MC simulations. The upper and lower figures show the IMC charge for events with $2 < E_{\rm TASC} < 6.3$~TeV and $6.3 < E_{\rm TASC} < 20$~TeV, respectively. Examples of charge distributions in the energy region below 2 TeV for CHD and IMC, and their correlation are shown in Fig.~S1 of the Supplemental Material~\cite{CALET2021-proton-SM}. 
}
    \label{fig:distZ}
  \end{center}
\end{figure}

A charge selection of proton candidates is performed 
by applying simultaneous window cuts on CHD and IMC reconstructed charges. 
Energy dependent criteria are defined for ``target'' events
 to maintain the same efficiency for the CHD and IMC.
In the higher energy region, the identification using IMC is useful to reject helium events.
Figure \ref{fig:distZ} shows examples of the $Z$ distribution using IMC.
Further details on the selection criteria can be found in the Supplemental Material~\cite{CALET2021-proton-SM} and in Ref.~\cite{CALET-p}.
\\
\indent
Background contamination is estimated using MC simulations of protons,
helium, and electrons.
Below $\sim5$~TeV (TASC energy deposits sum), the dominant background comes from off-acceptance protons.
The contamination is estimated below a few percent.
At higher energies, helium is the main background source, and the contamination gradually increases with the observed energy reaching a maximum of 20\% as shown in Fig. S2 of the Supplemental Material~\cite{CALET2021-proton-SM}.
A background contamination correction, based on the charge distribution, is applied 
before application of the energy unfolding.

The calorimetric energy resolution for protons is around 30\%--40\% with an observed energy fraction close to 35\%.
Therefore, energy unfolding is required to correct for bin-to-bin migration effects.
We follow a Bayesian approach, as implemented in the RooUnfold 
package~\cite{roounfold, BayesUnfolding} in ROOT~\cite{root}, 
whereby the response matrix is derived from the MC simulations. 
The unfolded energy spectrum is presented and compared with the $E_\text{TASC}$ distribution in Fig.~S3 of Supplemental Material~\cite{CALET2021-proton-SM}.
Convergence is usually reached within two iterations, 
given the relatively accurate prior distribution obtained from 
the previous observations, i.e., by AMS-02~\cite{AMS-02-proton} and 
CREAM-III~\cite{CREAM-III-pHe}.

The proton spectrum is obtained by 
correcting the effective geometrical acceptance with 
the unfolded energy distribution as follows:
\begin{eqnarray*}
 \Phi(E) & = & \frac{n(E)}{(S \Omega)_{\rm eff}(E) \,\, T \, \Delta E},\\
\label{fluxcalc}
n(E) & = & U\bigl(n_{\rm obs}(E_{\rm TASC}) - n_{\rm bg}(E_{\rm TASC})\bigr),
\label{fluxcalc2}
\end{eqnarray*}
where $\Delta E$ denotes the energy bin width, $U()$ the  
unfolding procedure operator based on the Bayes theorem, $n(E)$ the 
 bin counts of the unfolded distribution, $n_{\rm obs}(E_{\rm TASC})$ those 
of the observed energy distribution (including background), $n_{\rm bg}(E_{\rm TASC})$ the bin counts 
of background events in the observed energy distribution, $(S\Omega)_{\rm eff}$ 
the effective acceptance including all selection efficiencies, and $T$ the live time.\\
\indent
At the lowest energies, the HE-trigger efficiency drops significantly, and in this region LE-trigger events are
used instead. The event selection criteria for the HE and LE analyses are identical.
While the overall difference between the two selections is relatively small, the difference in the
low-energy region is sizeable while, in the energy region above 200 GeV, LE- and HE-trigger data are consistent.
Therefore we use LE-trigger data for $E<300$ GeV and HE-trigger data above.
The fluxes obtained with LE and HE triggers are presented within the respective energy regions in Fig.~S4 of Supplemental Material~\cite{CALET2021-proton-SM}.
\newline
\indent
$\bf{\it Systematic~uncertainties.-}$The systematic uncertainties include energy independent and dependent contributions.
The former is estimated around 4.1\% in total, from
the uncertainties on the live time (3.4\%), radiation environment
(1.8\%), and long-term stability (1.4\%).

\begin{figure}[h!]
\begin{center}
\includegraphics[width=1.0 \linewidth]{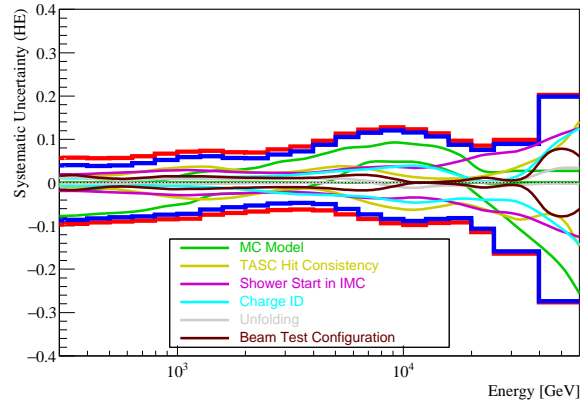}
\caption{Systematic uncertainties in the HE sample. The thick blue line shows the sum of the energy dependent systematic uncertainties. 
The thick red line is representative of the total systematic uncertainty, calculated as the quadratic sum of the various uncertainties, including the energy independent ones. A breakdown of the energy dependent uncertainties is also shown (thin internal lines). The systematic uncertainties of the HE sample are shown in an enlarged plot in Fig.~S5 of the Supplemental Material~\cite{CALET2021-proton-SM}.}
\label{fig:syst}
\end{center}
\end{figure}

\begin{figure}[h!] 
  \begin{center}
    \includegraphics[keepaspectratio,height=80mm,width=\hsize]{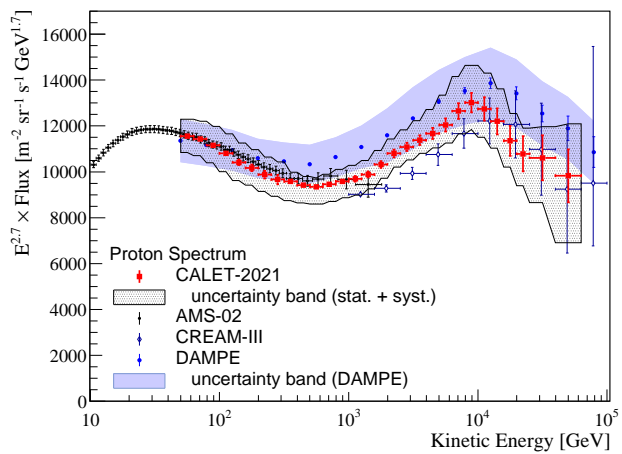}
  \end{center}
  \vspace{-0.5cm}
  \caption{Proton spectrum measured by CALET (red circles) compared with the experimental results of AMS-02 \cite{AMS-02-proton}, CREAM-III \cite{CREAM3_proton}, and DAMPE \cite{DAMPE_proton}. 
 The hatched band shows the total uncertainty for CALET as the quadratic sum of the various uncertainties. 
 The dark blue colored band shows the total uncertainty for DAMPE. The proton flux in tabular form can be found in the Supplemental Material~\cite{CALET2021-proton-SM}. }
  \label{fig:protonflux}
\end{figure}

The energy dependent component is estimated to be less than 10\% for $E<10$ TeV.
We take into account the uncertainties on MC model dependence,
IMC track consistency with the TASC energy deposits, shower start in the IMC, charge identification,
energy unfolding, and beam test configuration. For $E>10$ TeV the uncertainties
on MC model dependence and charge identification become dominant.
In the interval $10<E<40$ TeV the uncertainty is below 20\% while reaching
a maximum $\sim$ 30\% in the last bin.
Figure \ref{fig:syst} shows the systematic uncertainty in the HE
sample as a function of energy.
%
\newline
\indent
$\bf{\it Results.-}$Our extended measurement of the proton spectrum from 50 GeV to 60 TeV
is shown in Fig.~\ref{fig:protonflux}. The CALET flux is compared with AMS-02 \cite{AMS-02-proton}, DAMPE \cite{DAMPE_proton}, and CREAM-III \cite{CREAM3_proton}. 
Our spectrum is in good agreement with the rigidity spectra measured by magnetic spectrometers in the sub-TeV region, and it is also consistent, within the errors, with the measurements carried out with calorimetric instruments at higher energies.

Our data confirm the presence of a spectral hardening at a few hundred GeV as reported in our previous proton Letter \cite{CALET-p} with a higher significance of more than 20 sigma (statistical error). We also observe a spectral softening around 10 TeV.   
\begin{figure} [h!] 
  \begin{center}
 \includegraphics[keepaspectratio,height=70mm,width=\hsize]{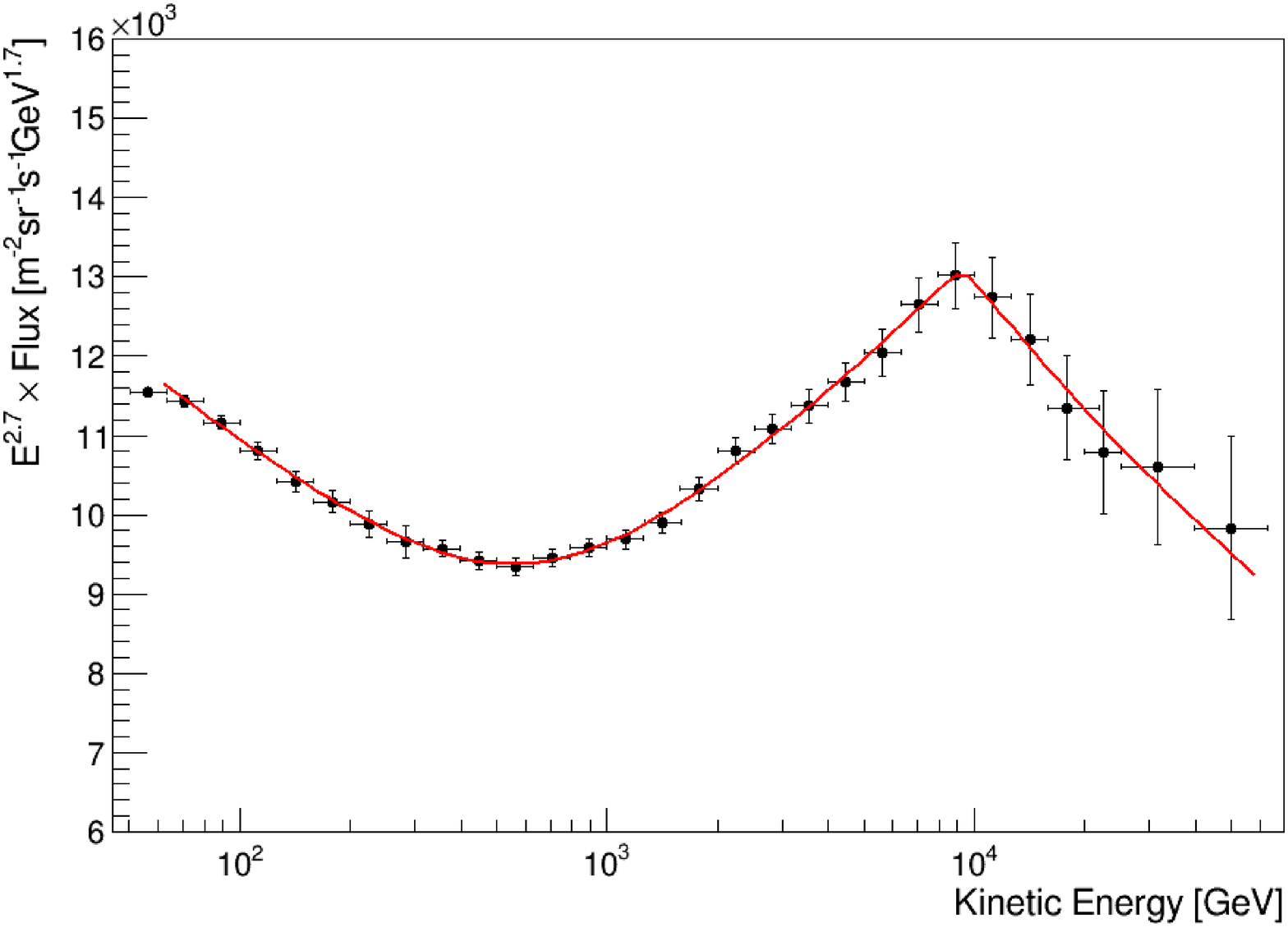}
  \end{center} 
  %
  \vspace{-0.5cm}
  \caption{A fit of the CALET proton spectrum (solid red line) with a double broken power law (Eq.\ref{eq:protonfit}). The horizontal error bars are representative of the bin width.}
  \label{fig:protonfit}
\end{figure}
%
We fit the proton spectrum in the energy region from 80 GeV to 60 TeV with a double broken power law (DBPL) function defined as follows:
\vspace{-0.1cm}
\begin{equation}
  \Phi^{\prime}(E) = E^{2.7}\times C\times\left(\frac{E}{1~\text{GeV}}\right)^\gamma\times \, \phi(E)
 \label{eq:protonfit}
\end{equation}
\vspace{-0.3cm}
with:
\begin{equation}
 \phi(E) =  \left[ 1+\left(\frac{E}{E_0}\right)^s\right]^{\frac{\Delta\gamma}{s}}\times \, \left[ 1+\left(\frac{E}{E_1}\right)^{s_{1}}\right]^{\frac{\Delta\gamma_1}{s_{1}}}
  \label{eq:protonfit2}
\end{equation}
where $\Phi^{\prime}(E)$ is the proton flux$\, \times \,E^{2.7}$, $C$ is a normalization factor, $\gamma$ the spectral index, $E_0$ is a characteristic energy of the region where a gradual spectral hardening is observed, $\Delta\gamma$ the spectral variation due to the spectral hardening, $E_1$ is a characteristic energy of the transition to the region of spectral softening, and $\Delta\gamma_1$ is the spectral index variation observed above $E_1$. Two independent smoothness parameters $s$ and $s_{1}$ are introduced in the energy intervals where spectral hardening and softening occur, respectively.
\begin{figure}[bth!]
\begin{center}
    \includegraphics[keepaspectratio,height=70mm,width=\hsize]{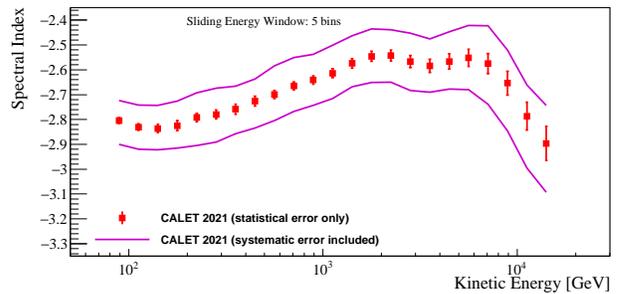}
    \caption{Energy dependence of the spectral index calculated within a sliding energy window for CALET (red squares). For each bin the spectral index is determined by fitting the data using $\pm$2 energy bins. The magenta curves indicate the uncertainty range including systematic errors.}
\label{fig:protonSpec}
\end{center}
\end{figure}
CALET data (black filled circles) and associated statistical errors are shown in Fig.~\ref{fig:protonfit} where the red line shows the best fitted function with parameters
%
%
$\gamma = -2.83 _{-0.02}^{+0.01}$, $s = 2.4_{-0.6}^{+0.8}$, $\Delta\gamma = 0.28_{-0.02}^{+0.04}$, $E_0 = 584 _{-58}^{+61}$ GeV, $\Delta\gamma_1 = -0.34_{-0.06}^{+0.06}$, $E_1 = 9.3_{-1.1}^{+1.4}$ TeV, and $s_{1} \sim$ 30 with a large error. 
The $\chi^2$ is 4.4 with 20 degrees of freedom.\\

\indent
Figure \ref{fig:protonSpec} shows the energy dependence of the spectral index calculated within 
a sliding energy window (red squares). The spectral index is determined for each bin by 
a fit to the data including the neighboring $\pm$2 bins in the region below 20 TeV
above which the highest two bins have relatively large errors.
Magenta curves indicate the uncertainty band including systematic errors.\\
\indent
As the hardening is very gradual, its onset (around 200 GeV) can be read off directly from Figure 5. It is followed by a sharp softening of the flux above$\,\sim{9}$ TeV.  The first spectral transition is found to be parameterized [Eq.(2)] by a relatively low value of $s$, while the second (sharper) one corresponds to a higher value of $s_{1}$ with a large uncertainty. Both parameters are left free in the fit. The fitted value of $E_0$ is found to be anticorrelated with the $s$ parameter. We additionally performed an independent fit to $\Delta\gamma$ and $\Delta\gamma_1$ with single-power-law functions in three energy sub-intervals, as shown in the Supplemental Material~\cite{CALET2021-proton-SM}. They were found to be consistent, within the errors, with the values obtained with the DBPL fit.
\vspace{+0.2cm}
\newline
\indent
$\bf{\it Conclusion.-}$We have measured the cosmic-ray proton spectrum covering 3 orders of magnitude in energy from 50~GeV to 60~TeV and characterized two spectral features in the high-energy CR proton flux with a single measurement in low earth orbit. Our new data extend the energy interval of our previous measurement \cite{CALET-p} while keeping a good consistency with our earlier result.
Our spectrum is not consistent with a single power law covering the whole range: (i) above a few hundred GeV we confirm our previous observation \cite{CALET-p} of a progressive spectral hardening, also reported by CREAM, PAMELA, AMS-02, and DAMPE; (ii) at energies around 10 TeV we observe a second spectral feature with a softening starting around 10 TeV.
In this energy region the shape of the spectrum  is consistent, within the errors, with the measurement reported by DAMPE. The results from two independent CALET analyses, with different efficiencies, were cross-checked and found in agreement.\\
\indent
Extended CALET operations were approved by JAXA/NASA/ASI in March 2021 through the end of 2024 (at least).
Improved statistics and refinement of the analysis, with additional data collected during the live time of the mission, will allow us to extend the proton measurement at higher energies and to reduce the systematic uncertainties.\\
\vspace{-0.3cm}
\newline
\indent
We gratefully acknowledge JAXA's contributions to the development of CALET and to the
operations onboard the International Space Station. We also 
express our sincere gratitude to ASI and NASA for
their support of the CALET project.
This work was supported in part by JSPS Grant-in-Aid for Scientific Research (S) Grant No.19H05608,
and by the MEXT-Supported Program for the Strategic Research Foundation at Private Universities
(2011--2015) (Grant  No.S1101021) at Waseda University. 
The CALET effort in Italy is supported by ASI under Agreement No. 2013-018-R.0 and its amendments.
The CALET effort in the U.S. is supported by NASA through Grants
No.  80NSSC20K0397, No. 80NSSC20K0399, and No. NNH18ZDA001N-APRA18-004.

\providecommand{\noopsort}[1]{}\providecommand{\singleletter}[1]{#1}  


\begin{thebibliography}{61}
\bibitem [{EPI()}]{EPICSver}   \BibitemOpen  \href@noop {} {\bibinfo  {journal} {EPICS and COSMOS versions are 9.20 and  8.00, respectively}\ }.
\end{thebibliography}
\end{document}


\widetext
\clearpage
\begin{center}
\end{center}
\setcounter{equation}{0}
\setcounter{figure}{0}
\setcounter{table}{0}
\setcounter{page}{1}
\makeatletter
\renewcommand{\theequation}{S\arabic{equation}}
\renewcommand{\thefigure}{S\arabic{figure}}
\renewcommand{\bibnumfmt}[1]{[S#1]}
\renewcommand{\citenumfont}[1]{S#1}

\begin{center}
\textbf{\Large Observation of Spectral Structures in the Flux of Cosmic-Ray Protons from 50~GeV to 60~TeV with Calorimetric Electron Telescope on the International Space Station}
\end{center}
\vspace*{0.5cm}
Supplemental material relative to "Observation of Spectral Structures in the Flux of Cosmic-Ray Protons from 50~GeV to 60~TeV with Calorimetric Electron Telescope on the International Space Station.''
\newpage
\begin{center}
{\bf DATA ANALYSIS}
\end{center}
The event selection and analysis are carried out as  in  our previous publication~\cite{CALET-p} with additional refinements and improvements. For instance, the charge selection with the IMC
  at high energies, which was not included in our previous publication,  is now implemented. In the following we mainly discuss the new analysis procedures that have been introduced in this paper.

\begin{center}
{\bf Charge Selection}
\end{center}

\begin{figure}[h!] 
\begin{center}
\includegraphics[width=16cm]{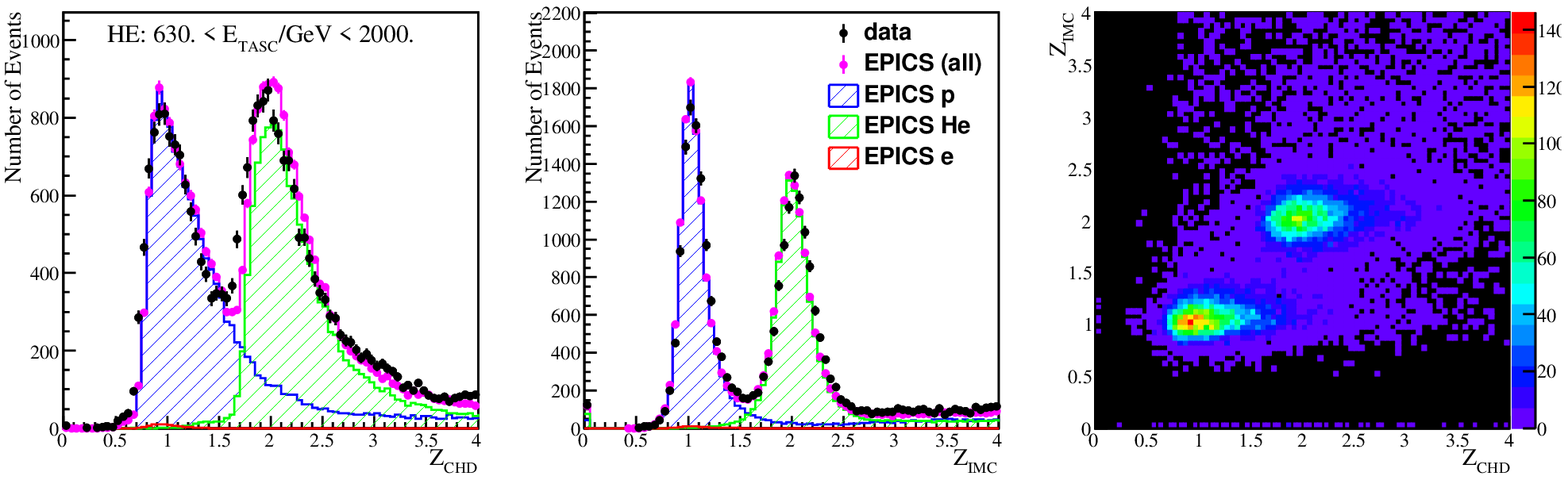}
\caption{Examples of charge distributions after applying criteria (1-7) (as explained in the main body of the paper) for CHD (left) and IMC (middle), and charge correlation plot (right) of IMC vs CHD. The events shown are selected with $630 < E_\text{TASC} < 2000$ GeV. The observed data are compared with the EPICS MC simulations.
}
\label{fig:distZ_SM}
\end{center}
\end{figure}

Above a few TeV the CHD charge measurement is affected by a copious background generated by backscattered radiation and integrated along the length of the scintillator paddles and across their relatively wide lateral segmentation (2 cm). Instead, as can be seen in the examples of charge distributions given in Figure~\ref{fig:distZ_SM}, below 2 TeV the background level is lower and we observe an excellent correlation of the IMC and CHD charge measurements. Above a few TeV, we take advantage of the high segmentation (1 mm) of the IMC fibers and apply charge cuts based only on the dE/dx information from the IMC fibers.  In the main body of the paper, Figure 1 shows an example of the effectiveness of the IMC charge cut which is carried out by applying an energy-dependent selection, optimized via MC simulations, to give a constant efficiency of $\sim 98\%$.  Charge selection based on the CHD signals uses a similar energy-dependent charge cut, albeit with a lower efficiency (close to $95\%$).


\begin{figure}[h!] 
\begin{center}
\includegraphics[width=8cm]{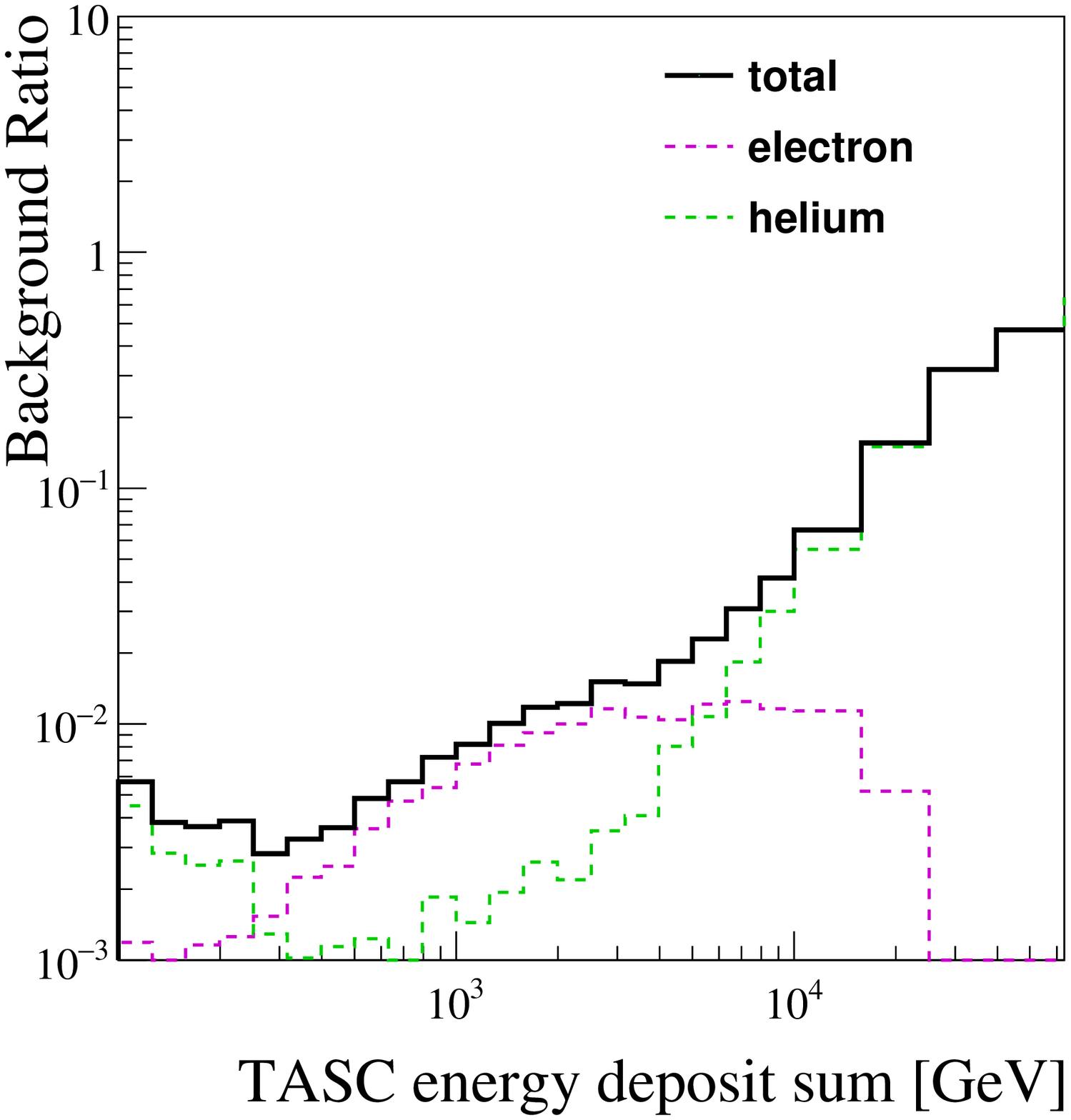}
\caption{Energy dependence of the background contamination due to the electron and helium contamination after subtracting off-acceptance protons (HE-trigger analysis). }
\label{fig:distBG_SM}
\end{center}
\end{figure}

\begin{figure}[h!]
\begin{center}
\includegraphics[width=0.45\hsize]{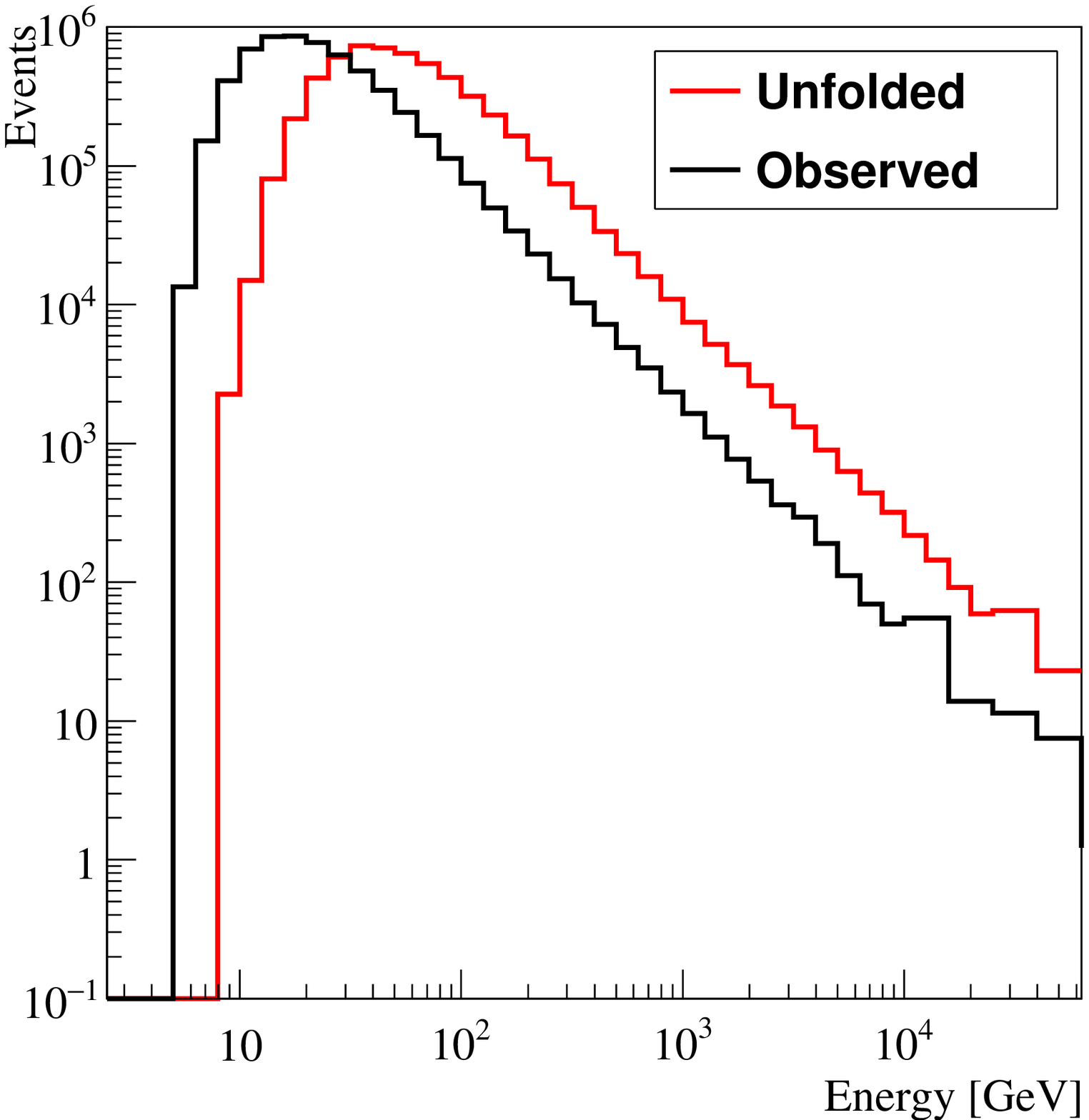}
\caption{Distribution of TASC energy deposits (Observed) after subtraction of the background contamination and the energy spectrum after the unfolding (Unfolded) in the HE-trigger analysis.}
\label{fig:distUnfolE_SM}
\end{center}
\end{figure}

\begin{figure}[h!] 
\begin{center}
\includegraphics[width=0.6\hsize]{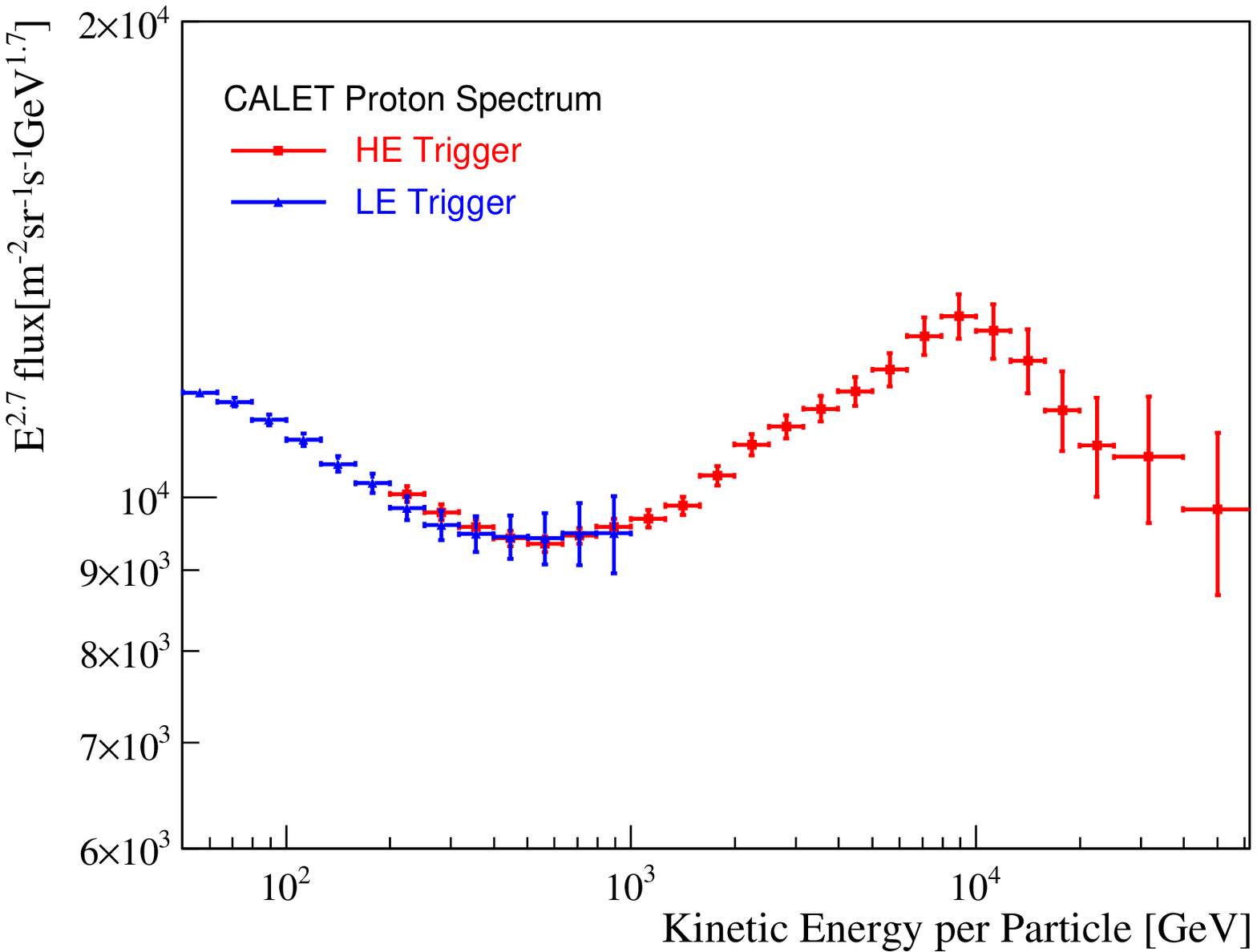}
\caption{Comparison of proton spectrum derived from two data sets corresponding to
  the HE-trigger (red) and LE-trigger (blue) analyses.}
\label{fig:distLEHE_SM}
\end{center}
\end{figure}

\begin{center}
{\bf Background Contamination}
\end{center}
Background contamination is estimated from the MC simulations of proton, helium and electron events as a function of the observed energy by assuming their spectral shape as derived from previous measurements,  i.e., AMS-02~\cite{AMS-02-proton, AMS-02-helium} and CREAM-III~\cite{CREAM-III-pHe}. The dominant component of the contamination comes from off-acceptance protons except for the highest energy region above 5~TeV.  For events taken with the HE-trigger, Figure~\ref{fig:distBG_SM} shows the background contamination, as a function of $E_\text{TASC}$, due to electron and helium events after removing the background from off-acceptance protons. 
The latter are efficiently removed by comparing the shower axis, as reconstructed in the TASC, with the direction of the track from the IMC.
As can be seen in the figure, above 5~TeV the contamination is dominated by helium and its contribution ranges from a few~\% to 20\% with a maximum in the highest energy region of $E_\text{TASC}\sim 20$~TeV corresponding to the primary proton energy of $\sim$60~TeV. The effect of backscattering becomes increasingly more significant in this energy region. In the lower energy region covered by the LE-trigger, the total contamination, inclusive of off-acceptance protons,  is suppressed and is lower than a few \%.  The larger contamination by helium above 5 TeV causes a significant contribution to the systematic errors as shown in Figure~\ref{fig:distBG_SM}. 
\begin{center}
{\bf Energy Unfolding}
\end{center}
Energy unfolding is adopted after subtraction of the background contamination as mentioned above. In this analysis, we use an unfolding method based on the Bayesian approach implemented in the RooUnfold package~\cite{roounfold, BayesUnfolding} in ROOT~\cite{root}, with the response matrix derived
from MC simulations. Convergence is obtained within two iterations, given the relatively accurate prior distribution
obtained from the previous observations, i.e., by AMS-02~\cite{AMS-02-proton}  and CREAM-III~\cite{CREAM-III-pHe}.  The energy spectrum obtained with the unfolding procedure is presented in Figure~\ref{fig:distUnfolE_SM} and compared
with the differential distribution of the TASC energy deposits as they were observed before the unfolding.  
\newpage
\begin{center}
{\bf Consistency between LE and HE Analyses}
\end{center}
The HE-trigger mode is always active on orbit to maximize the efficiency for high energy electrons above 10~GeV.
For the data analysis, we applied an additional offline-trigger with a higher threshold to select events while avoiding the effects due to gain variations during the long observation period. Since the offline trigger threshold is higher than the one of the hardware trigger, the HE-trigger analysis introduces an efficiency bias at energies below 200 GeV, which is studied with a scan of the offline-trigger threshold using LE-trigger data. Therefore, LE-Trigger analysis is required below 200~GeV.  As presented in Figure~\ref{fig:distLEHE_SM}, the two spectra obtained with the LE and HE analyses are very well consistent in the energy region shared by the two analyses. They are combined around 300 GeV, taking into account the
different statistics of the two trigger modes. As a result, the whole  energy range from 50 GeV to 60 TeV has been observed by CALET.

\begin{figure}[h!] 
\begin{center}
\includegraphics[width=0.8\hsize]{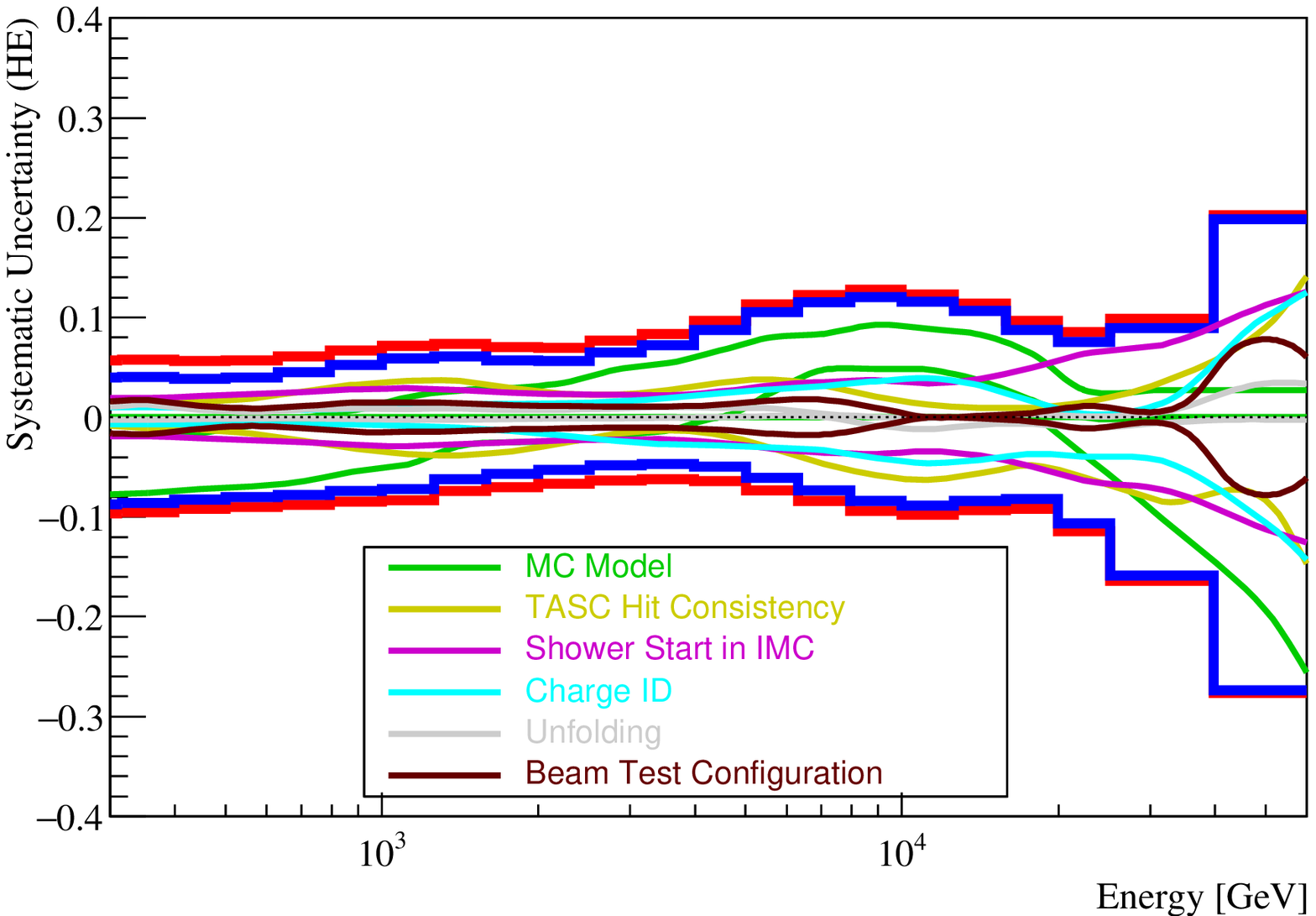}
\caption{Systematic uncertainties for the HE-trigger sample.}
\label{fig:distSys_SM}
\end{center}
\end{figure}

\begin{center}
{\bf Systematic Uncertainties}
\end{center}
%
Systematic uncertainties are estimated with LE and HE samples, respectively and shown in Figure~\ref{fig:distSys_SM} for the HE sample.
Energy-independent systematic uncertainties are estimated around 4.1\% in total, from the uncertainties on the live time (3.4\%), long-term stability (1.4\%), and radiation environment (1.8\%). The latter contribution is mainly due to effects due to the ISS radiation environment. No time-dependent component due to the aging of the instrument is included in this category.  Since low energy radiation rapidly increases in the polar region, these effects can be evaluated using the rigidity cutoff dependence of the observed spectrum. To study this uncertainty, flight data were divided into 4 cases of rigidity cutoff range as (i) less than 2GV, (ii) 2-6GV, (iii) 6-10GV, and (iv) above 10GV. The systematic error due to ISS radiation environment was evaluated in the 20-100 GeV energy range resulting in a systematic uncertainty of 1.8\%. 

In addition to energy independent uncertainties, several sources of systematics are taken into account and shown separately, as a function of energy, in the same figure. They are denoted with the following labels: (1) "MC Model" for the dependence due to the uncertainties on the current models of hadronic interactions, (2) "TASC Hit Consistency", (3) "Shower Start in IMC", (4)  "Charge ID" for the charge identification, (5) "Unfolding" for the energy response, and (6) "Beam Test Configuration". The uncertainty on the hadronic interaction models is estimated from the comparison of EPICS with GEANT4 and FLUKA. Contributions (2), (3), and (4) come from uncertainties on the event reconstruction. The uncertainty (5), stemming from the energy response, is estimated by summing up the energy deposits closest to the shower axis. 
We extrapolate the reconstruct track and determine its impact point on the TASC. By comparing different choices in the list of neighbor logs that we include in the sums of deposits, we estimate the uncertainties associated to the measurement of the energy sum in the TASC due to the calorimeter granularity and to the shower width.
An additional uncertainty (6) is estimated from the study of the difference between flight and beam test configurations. Details can be found in the Supplemental Material of~\cite{CALET-p}. For $E<10$~TeV, the overall uncertainty is less than 10\%. For $E>10$~TeV, the uncertainties from the MC model dependence and charge identification become large, reaching 30\% at maximum.\\

\clearpage
\begin{center}
{\bf Results}
\end{center}

\begin{figure}[h!] 
\begin{center}
\includegraphics[width=16cm]{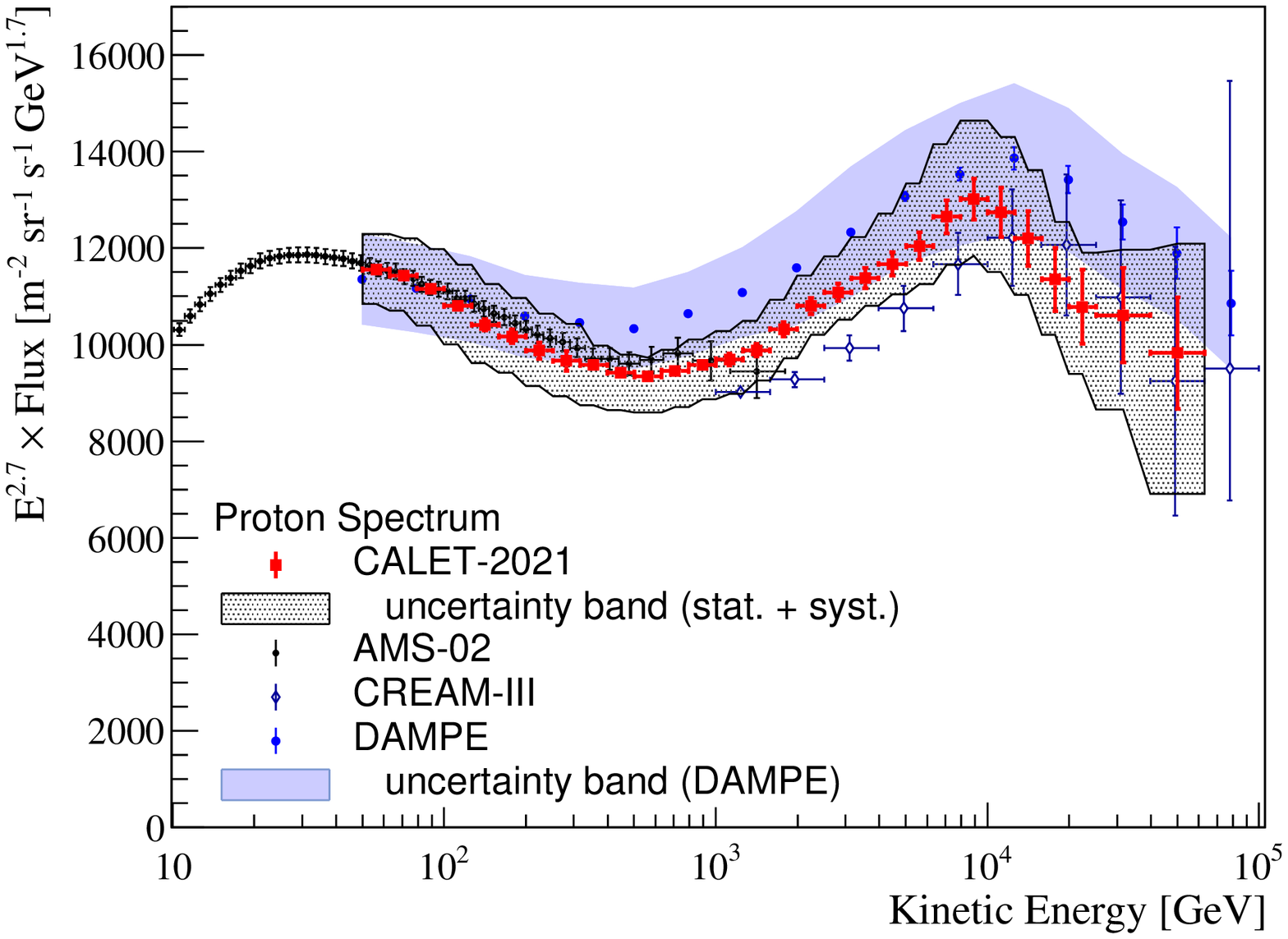}
\caption{Cosmic-ray proton spectrum observed with CALET from 50~GeV to~60 TeV using an energy binning with 10 bins per decade, except for the last two (wider) bins. The horizontal error bars are representative of the bin widths.
The hatched band indicates the total uncertainty as the quadratic sum of systematic and statistical errors. Direct measurements in space by AMS-02~\cite{AMS-02-proton},  DAMPE~\cite{DAMPE-proton} and by balloon observation with CREAM~\cite{CREAM-III-pHe} are included for comparison. The dark blue band shows the total uncertainty for DAMPE. }
\label{fig:CALET_Spectrum_SM}
\end{center}
\end{figure}

The proton spectrum measured by CALET (red circles) is shown in Figure~\ref{fig:CALET_Spectrum_SM} where it is compared with the experimental results of AMS-02, CREAM-III, and DAMPE. The hatched band shows the total uncertainty for CALET. The dark blue colored band shows the total uncertainty for DAMPE.

\clearpage
\begin{center}
{\bf Additional fit of the proton spectrum in energy sub-intervals}
\end{center}

\begin{figure}[h!] 
\begin{center}
\includegraphics[width=0.8\hsize]{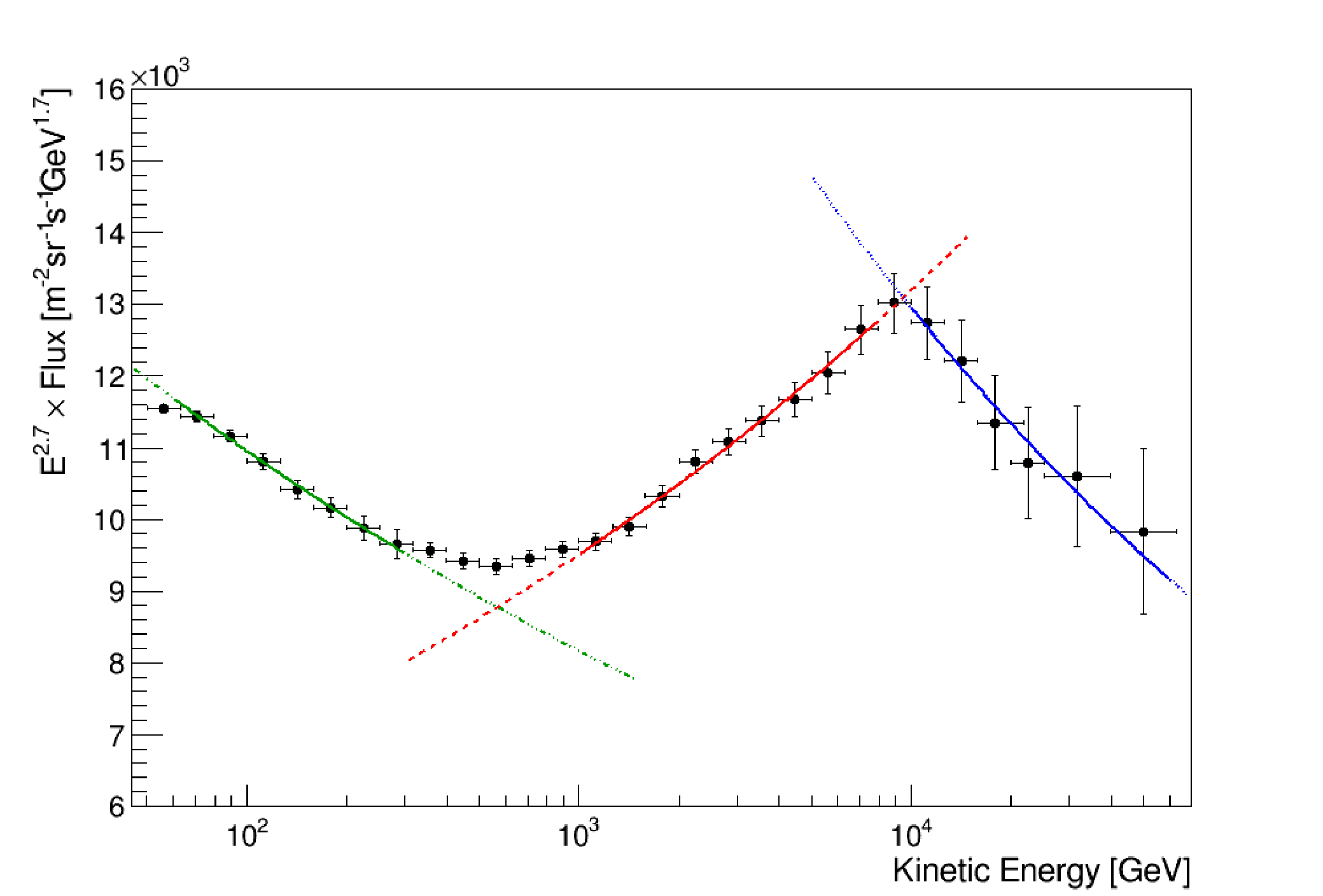}

\caption{Fits of CALET proton spectrum with single power-law functions. Green, red, and blue (solid) lines show the best fitted results in the energy intervals 60 -- 300 GeV, 1 -- 8 TeV, and 10 -- 60 TeV. The corresponding fitted indices are $-2.83\pm 0.02$, $-2.56\pm 0.01$, and $-2.89\pm 0.07$, respectively. The values of $\Delta \gamma$ and $\Delta \gamma_{1}$ are consistent, within the errors, with the values obtained with the DBPL fit, as described in the main section of the paper.
 }

\label{fig:distFitSPL}
\end{center}
\end{figure}

Separate fits of the proton spectrum, with single power-law functions, are shown in three different energy sub-intervals in Figure~\ref{fig:distFitSPL}  where the error bars refer to purely statistical errors.  

\renewcommand{\arraystretch}{1.25}
\begin{table*}[hbtp!]
\caption{
CALET proton spectrum in tabular form. A representative energy value is evaluated for each bin from the geometric mean of the lower and upper edges. 
The first and second error associated with the bin energy represent 
the systematic error on the global energy scale and an energy dependent
error, respectively. For the flux, the first, second and third
error represent the statistical uncertainty (68\% confidence level),
a systematic uncertainty on the normalization, and an energy dependent 
systematic uncertainty, respectively.
\label{tab:pro}}
\begin{ruledtabular}
\begin{tabular}{cccrrrrrrr}
Energy Bin & Representative Energy & Flux \\
(GeV)  & (GeV) & (m$^{-2}$sr$^{-1}$s$^{-1}$GeV$^{-1}$) \\ 
\hline
$(5.01 - 6.31) \times 10^{1}$ & $(5.62 \pm 0.16 \pm 0.06)\times 10^{1}$ & $(2.18 \pm 0.01 \pm 0.09 _{-0.10}^{+0.10})\times 10^{-1}$\\
$(6.31 - 7.94) \times 10^{1}$ & $(7.08 \pm 0.20 \pm 0.07)\times 10^{1}$ & $(1.16 \pm 0.01 \pm 0.05 _{-0.06}^{+0.06})\times 10^{-1}$\\
$(7.94 - 10.00) \times 10^{1}$ & $(8.91 \pm 0.25 \pm 0.09)\times 10^{1}$ & $(6.06 \pm 0.05 \pm 0.25 _{-0.33}^{+0.36})\times 10^{-2}$\\
$(1.00 - 1.26) \times 10^{2}$ & $(1.12 \pm 0.03 \pm 0.01)\times 10^{2}$ & $(3.15 \pm 0.03 \pm 0.13 _{-0.19}^{+0.21})\times 10^{-2}$\\
$(1.26 - 1.59) \times 10^{2}$ & $(1.41 \pm 0.04 \pm 0.01)\times 10^{2}$ & $(1.63 \pm 0.02 \pm 0.07 _{-0.10}^{+0.11})\times 10^{-2}$\\
$(1.59 - 2.00) \times 10^{2}$ & $(1.78 \pm 0.05 \pm 0.02)\times 10^{2}$ & $(8.56 \pm 0.12 \pm 0.35 _{-0.50}^{+0.55})\times 10^{-3}$\\
$(2.00 - 2.51) \times 10^{2}$ & $(2.24 \pm 0.06 \pm 0.02)\times 10^{2}$ & $(4.46 \pm 0.08 \pm 0.18 _{-0.25}^{+0.27})\times 10^{-3}$\\
$(2.51 - 3.16) \times 10^{2}$ & $(2.82 \pm 0.08 \pm 0.03)\times 10^{2}$ & $(2.35 \pm 0.05 \pm 0.10 _{-0.13}^{+0.14})\times 10^{-3}$\\
$(3.16 - 3.98) \times 10^{2}$ & $(3.55 \pm 0.10 \pm 0.04)\times 10^{2}$ & $(1.25 \pm 0.01 \pm 0.05 _{-0.11}^{+0.05})\times 10^{-3}$\\
$(3.98 - 5.01) \times 10^{2}$ & $(4.47 \pm 0.13 \pm 0.05)\times 10^{2}$ & $(6.59 \pm 0.08 \pm 0.27 _{-0.54}^{+0.25})\times 10^{-4}$\\
$(5.01 - 6.31) \times 10^{2}$ & $(5.62 \pm 0.16 \pm 0.06)\times 10^{2}$ & $(3.51 \pm 0.04 \pm 0.14 _{-0.28}^{+0.14})\times 10^{-4}$\\
$(6.31 - 7.94) \times 10^{2}$ & $(7.08 \pm 0.20 \pm 0.07)\times 10^{2}$ & $(1.91 \pm 0.02 \pm 0.08 _{-0.15}^{+0.09})\times 10^{-4}$\\
$(7.94 - 10.00) \times 10^{2}$ & $(8.91 \pm 0.25 \pm 0.10)\times 10^{2}$ & $(1.04 \pm 0.01 \pm 0.04 _{-0.08}^{+0.05})\times 10^{-4}$\\
$(1.00 - 1.26) \times 10^{3}$ & $(1.12 \pm 0.03 \pm 0.01)\times 10^{3}$ & $(5.64 \pm 0.07 \pm 0.23 _{-0.41}^{+0.33})\times 10^{-5}$\\
$(1.26 - 1.59) \times 10^{3}$ & $(1.41 \pm 0.04 \pm 0.02)\times 10^{3}$ & $(3.09 \pm 0.04 \pm 0.13 _{-0.19}^{+0.19})\times 10^{-5}$\\
$(1.59 - 2.00) \times 10^{3}$ & $(1.78 \pm 0.05 \pm 0.03)\times 10^{3}$ & $(1.73 \pm 0.02 \pm 0.07 _{-0.10}^{+0.10})\times 10^{-5}$\\
$(2.00 - 2.51) \times 10^{3}$ & $(2.24 \pm 0.06 \pm 0.04)\times 10^{3}$ & $(9.74 \pm 0.15 \pm 0.40 _{-0.51}^{+0.55})\times 10^{-6}$\\
$(2.51 - 3.16) \times 10^{3}$ & $(2.82 \pm 0.08 \pm 0.05)\times 10^{3}$ & $(5.37 \pm 0.09 \pm 0.22 _{-0.26}^{+0.35})\times 10^{-6}$\\
$(3.16 - 3.98) \times 10^{3}$ & $(3.55 \pm 0.10 \pm 0.06)\times 10^{3}$ & $(2.96 \pm 0.06 \pm 0.12 _{-0.14}^{+0.21})\times 10^{-6}$\\
$(3.98 - 5.01) \times 10^{3}$ & $(4.47 \pm 0.13 \pm 0.08)\times 10^{3}$ & $(1.63 \pm 0.03 \pm 0.07 _{-0.08}^{+0.14})\times 10^{-6}$\\
$(5.01 - 6.31) \times 10^{3}$ & $(5.62 \pm 0.16 \pm 0.10)\times 10^{3}$ & $(9.03 \pm 0.22 \pm 0.37 _{-0.55}^{+0.95})\times 10^{-7}$\\
$(6.31 - 7.94) \times 10^{3}$ & $(7.08 \pm 0.20 \pm 0.13)\times 10^{3}$ & $(5.09 \pm 0.14 \pm 0.21 _{-0.37}^{+0.59})\times 10^{-7}$\\
$(7.94 - 10.00) \times 10^{3}$ & $(8.91 \pm 0.25 \pm 0.17)\times 10^{3}$ & $(2.81 \pm 0.09 \pm 0.12 _{-0.24}^{+0.34})\times 10^{-7}$\\
$(1.00 - 1.26) \times 10^{4}$ & $(1.12 \pm 0.03 \pm 0.02)\times 10^{4}$ & $(1.48 \pm 0.06 \pm 0.06 _{-0.13}^{+0.17})\times 10^{-7}$\\
$(1.26 - 1.59) \times 10^{4}$ & $(1.41 \pm 0.04 \pm 0.03)\times 10^{4}$ & $(7.61 \pm 0.36 \pm 0.31 _{-0.64}^{+0.81})\times 10^{-8}$\\
$(1.59 - 2.00) \times 10^{4}$ & $(1.78 \pm 0.05 \pm 0.04)\times 10^{4}$ & $(3.80 \pm 0.22 \pm 0.16 _{-0.31}^{+0.33})\times 10^{-8}$\\
$(2.00 - 2.51) \times 10^{4}$ & $(2.24 \pm 0.06 \pm 0.05)\times 10^{4}$ & $(1.94 \pm 0.14 \pm 0.08 _{-0.21}^{+0.15})\times 10^{-8}$\\
$(2.51 - 3.98) \times 10^{4}$ & $(3.16 \pm 0.09 \pm 0.06)\times 10^{4}$ & $(7.51 \pm 0.69 \pm 0.31 _{-1.19}^{+0.67})\times 10^{-9}$\\
$(3.98 - 6.31) \times 10^{4}$ & $(5.01 \pm 0.14 \pm 0.10)\times 10^{4}$ & $(2.01 \pm 0.24 \pm 0.08 _{-0.55}^{+0.40})\times 10^{-9}$\\

\end{tabular}
\end{ruledtabular}
\end{table*}

\clearpage

\providecommand{\noopsort}[1]{}\providecommand{\singleletter}[1]{#1}